# Porous LSCF/Dense 3YSZ Interface Fracture Toughness Measured by Single Cantilever Beam Wedge Test

Xin Wang[a], Fei He[b], Zhangwei Chen[a] and Alan Atkinson[a]

[a] Department of Materials, Imperial College London SW7 2BP, UK
[b] Key Laboratory of Science and Technology on Advanced Composites in Special Environments, Harbin Institute of Technology, P.R. China

**Abstract**

Sandwich specimens were prepared by firing a thin inter-layer of porous $La_{0.6}Sr_{0.4}Co_{0.2}Fe_{0.8}O_3$ (LSCF) to bond a thin tetragonal yttria-stabilised zirconia (YSZ) beam to a thick YSZ substrate. Fracture of the joint was evaluated by introducing a wedge between the two YSZ adherands so that the stored energy in the thin YSZ cantilever beam drives a stable crack in the adhesive bond and allows the critical energy release rate for crack extension (fracture toughness) to be measured. The crack path in most specimens showed a mixture of adhesive failure (at the YSZ/LSCF interface) and cohesive failure (within the LSCF). It was found that the extent of adhesive fracture increased with firing temperature and decreased with LSCF layer thickness. The adhesive failures were mainly at the interface between the LSCF and the thin YSZ beam and FEM modelling revealed that this is due to asymmetric stresses in the LSCF. Within the firing temperature range of 1000-1150 ºC, the bonding fracture toughness appears to have a strong dependence on firing temperature. However, the intrinsic adhesive fracture toughness of the LSCF/YSZ interface was estimated to be 11 J m$^{-2}$ and was not firing temperature dependent within the temperature range investigated.

## I Introduction

The development of solid oxide fuel cells (SOFCs) is typically guided by progress in electrochemical performance, but with the utilization of SOFCs in the larger power unit of a stack, mechanical aspects are also receiving rising interest [1]. A planar SOFC cell consists of three basic layers (anode, electrolyte, and cathode). The materials are rigidly bonded in the multilayer structure and differences in materials properties result in residual stresses. Such stresses can arise from the co-firing of the cells, differences in thermal-expansion coefficients, thermal gradients, and chemical gradients of the diffusing species. Additional stresses can also be introduced from the final arrangement and fixation of the cells in the SOFC stack [2].

Interface fracture energy is a parameter that is key to evaluating the robustness of multilayer systems developed in various technological applications. In the case of SOFCs where two porous electrodes are positioned on each side of a dense ceramic electrolyte [3]. The electrode/electrolyte interface has to withstand mechanical stresses that arise during fabrication and operation. Interface damage, even if not catastrophic, often results in poor electrical contact and degradation of electrochemical performance. Therefore, interface fracture toughness is very important in the assessment of the mechanical reliability of laminated structures such as the design of robust SOFCs with a long lifetime.

LSCF represents a family of perovskite-structured materials with general formula $La_{1-x}Sr_xCo_yFe_{1-y}O_{3-\delta}$ that are good candidates for cathode materials for SOFCs, due to their promising mixed electronic-ionic conductivity and high oxygen surface exchange rate [2-5]. Both 3YSZ and 8YSZ (zirconia containing 3mol% and 8mol% $Y_2O_3$ respectively) are favourable electrolyte materials for SOFCs. 8YSZ has the higher ionic conductivity, but 3YSZ has higher mechanical strength and toughness. In this paper we have chosen to investigate the interface between porous LSCF and dense 3YSZ as an example.

Several different test methods have been proposed for measuring interface fracture toughness: such as double cantilever beam [6]; four point bending [7]; double cleavage drilled compression [5]; indentation [8, 9]; wedge impression [10]; and cross sectional indentation [11-14]. However, these are not well-suited to investigate a porous ceramic film on a dense ceramic substrate. Indentation or impression methods rely on substantial plastic deformation of the substrate material to provide the driving force for interface crack propagation and are therefore not suitable for an all-ceramic system. The double cleavage drilled compression [5] method allows the interface toughness to be measured even when it exceeds the fracture toughness of the adjoining materials because of the stabilising role of the compressive loads [4]. However, specimen preparation for an all-ceramic system can be very expensive, if not impossible. Four point bending of a notched laminate beam [7] is appropriate when the fracture toughnesses of the materials involved are sufficiently high to prevent vertical cracking, which would not be the case for a system containing porous ceramics. Vertical cracking and/or segmentation can readily occur in porous materials and would decrease the stored elastic energy in the laminate and make the evaluation of the interface fracture energy unreliable. A further restriction of this method is a limit on the debonding layer thickness. There exists a critical thickness to store sufficient energy for crack propagation at the debonding interface [15]. Consequently, Hofinger et al [15] modified the original method by adding a stiffening layer to prevent vertical cracking and segmentation and provide sufficient driving force for interface crack propagation. This modified 4 point bending method has been successfully used to measure the interface fracture energy for a porous composite cathode on a YSZ electrolyte [3] and the interface between current collector and sealant in multilayered cells [2]. Sørenson and Horsewell [16] employed a special test fixture which loads a double cantilever beam sandwich specimen with pure bending moments and provides stable crack growth. Crack growth was detected by in situ SEM observation. The macroscopic fracture energy of the interface between dense lanthanum strontium chromite and a porous lanthanum strontium manganite was measured to lie in the range of 1.4 – 3.8 J/m$^2$ [16].

In the current work, a wedged single cantilever beam method, with a long thin beam, was designed to measure the fracture energy of the joint between a porous ceramic film and a dense ceramic substrate. In this method, the decrease of the stored energy in the system associated with cracking and/or segmentation of the porous material is negligible. There is always sufficient driving force to propagate fracture because the thickness of the beam can be changed according to the requirement. Another advantage of this method is that the accuracy of the fracture energy release rate is not sensitive to the exact position of the crack tip, which avoids the necessity of using in-situ high resolution microscopy to determine the crack tip position.

## II Experimental

### 1) Specimen preparation

Specimens were made in which thin 3YSZ beams (either 50x5x0.3 mm or 50x5x0.15 mm) were bonded to thick 3YSZ substrates (10mm in thickness) using a porous LSCF film as the adhesive. The LSCF sandwiched between the beam and the substrate was applied in the form of a slurry (ink). After drying and firing, the LSCF became a thin porous solid layer with a thickness of 10-30 μm.

3YSZ plates were supplied by Kerafol GmbH (Eschenbach, Germany), and a LSCF screen-printing ink (LSCF6428) was provided by ESL-UK. The ink was modified by diluting the original ink with terpineol (Sigma, UK) at a volume ratio of ink to terpineol of 1:2 and then homogenized by ball milling. In earlier work it was found that films fabricated using the as-received ink tended to have cracks, whereas cracking was avoided using the modified ink [17]. Therefore in the current work all the specimens were prepared using the modified ink. The sintering of the sandwiched layer would be subject to a similar constraint as an electrolyte film except for near the edges of the film. Since this extends over a lateral distance from the edge equal to a few times the film thickness, this is negligible as compared to the total bonded area ( ~ 25mmx5mm) .

Two different methods were used to apply the LSCF: either as a single layer (denoted as SL hereafter) or as a triple layer (TL hereafter). In the SL method, a single layer of wet LSCF film was first tape-cast on the substrate using a mask, then the 3YSZ beam was placed on the top of the wet LSCF film. Before LSCF ink application, the substrates were ground using grade 120 silicon carbide paper and the thin beams were used in their as-received state. Both substrate and thin beams were carefully cleansed with acetone before applying LSCF ink. Assemblies were dried for 12 hours at 100 °C and then sintered at different temperatures. In the TL method, LSCF films were first tape-cast on both the substrate and the thin beam. After drying ( at 100 °C for 24 h), a third layer of LSCF ink was applied on top of the dried LSCF film on the substrate, and then the thin beam (already coated with a dried film of LSCF) was carefully placed on the wet LSCF third layer. The 3-layered assembly was dried again at 100 °C for 24 h and then sintered at different temperatures. For the firing, a rate of 5 °C /min was used for both heating and cooling, the dwell time at the top temperature was 2 hrs.  A load of 50 grams (an alumina block) was placed on the assembly during drying and firing in order to maintain contact between components and enhance bonding.

## 2) Single beam wedge test

Fig. 1a shows a specimen under test in the process of delamination and Fig.1b is a schematic diagram of the test arrangement. The YSZ beam thickness is $t_0$ (either 0.15 or 0.3 mm) and the LSCF bonding layer thickness $t_1$ is 10~30μm, and is on the right hand part of the assembly in these figures. The shortest cantilever beam length $L_{Bo}$ (at the beginning of the test, before any crack propagation in the joint) was about 20 mm.

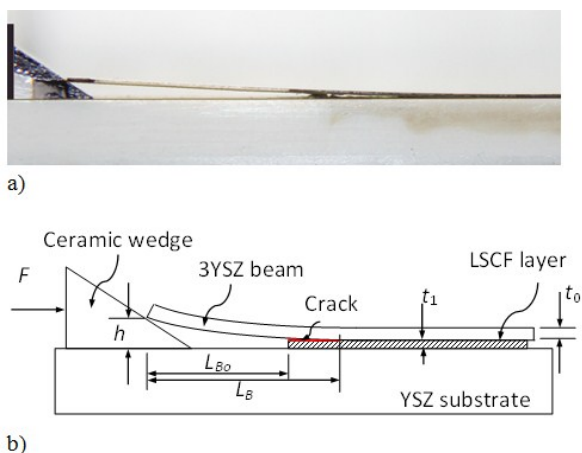

*Fig.1 a) An optical image of a specimen under test; b) a schematic of the single beam wedge test.*

During the test, an advancing ceramic (3YSZ) wedge generates a vertical displacement (***h***) of the beam at the left, provoking fracture of the joint between the beam and the substrate. The wedge slides on the YSZ substrate and has a tip angle of 30º. Its contact with both the substrate and beam was lubricated with graphite to reduce friction.

At each step, a small wedge advancement (approximately 0.1mm) was used to increase the stored energy in the bending beam, which is the driving force for fracture of the joint. In turn propagation of the fracture along the joint decreases the stored energy in the system. Therefore the crack in the joint reaches a new position of equilibrium and becomes stable. Since the layer thickness of the porous LSCF is much thinner than the dense YSZ beam, and its elastic modulus is much lower, the stored energy in the bending beam accounts for more than 99% of the total stored energy in the system.  Therefore, the stored energy released by crack propagation in the joint is independent of the failure type (either an adhesive failure at the upper or lower LSCF/YSZ interface or a cohesive failure within the LSCF). Furthermore, any stored elastic energy release associated with the vertical fracture and segmentation of the porous LSCF is negligible compared with the energy released by failure propagating along the joint.

We next evaluate the energy release rate in the bent YSZ beam as a function of the crack propagation along the joint. The stored elastic energy in a cantilever beam (the thin YSZ in this case) with a concentrated load at the end is given by [18]:

$$U = \frac{3EIh^2}{2L_B^3} \qquad \text{Eq. 1}$$

where ***E*** is the Young's modulus of the beam, $L_B$ the cantilever length and ***h*** is the beam deflection at the loading point. ***I*** is the second moment of area: ***I=$bt_0^3$/12***, with ***b*** and $t_0$ being the width and thickness of the beam.

Interface fracture propagation to the right in Figure 1b leads to an increase of the cantilever length. In a stable process, an increase in ***h*** (***δh***) would generate an increase of store energy (***δU***), which can be released by an increase of the cantilever length $δL_B$. Therefore the critical energy release rate ***Gc*** for crack propagation along the joint can be expressed:

$$G_C = -\frac{1}{b}\frac{dU}{dL_B} = \frac{9EIh^2}{2bL_B^4} \qquad \text{Eq. 2}$$

In Eq.2, the Young's modulus for 3YSZ: ***E*** = 200 GPa [19, 20]. The second moment of area ***I*** and beam width **b** are fixed, whereas the other two remaining parameters: **h** and $L_B$ are changing during the test.

According to Eq.2, $G_c \propto L_B^{-4}$, so accurate measurement of beam length is important for accurate evaluation of interface fracture energy. The beam length is equal to the original beam length plus the length of the crack in the joint and therefore in principle it is necessary to know the position of the crack tip. However, because the original beam length is large, the sensitivity to the crack tip position is reduced. The error tolerance for crack tip determination is: $\delta C = L_B \cdot e_{Gc}/4$, where $e_{Gc}$ is the acceptable fractional error for **Gc**. Thus a greater $L_B$ allows a large δC to be tolerated. In the current work, $L_B$ >20mm. If we assume an error of 10% is acceptable for **Gc**, then the permissible error in **δC** is estimated to be >0.5mm. Hence macroscopic optical imaging is sufficient for determining the crack tip position.

In the present experiments, a high resolution picture (e.g. Fig.1a) was taken after every step of wedge advancement, with a resolution of 0.017mm/pixel and the crack tip was located using a beam profile fitting method. Engauge Digitizer, an open source digitizing software, was used to convert the beam profile image into (x,y) coordinates. The digitized beam profile was then fitted to the theoretical profile, namely:

$$y = \frac{P}{6EI}(3x^2 L - x^3) \quad \text{Eq. 3}$$

More details about beam profile fitting method can be found in references [21, 22]. This is a more reliable way of determining the crack tip position because it makes use of the full beam profile.

During the test, the cantilever beam length $L_B$ would gradually increase from ~20 to ~45 mm as the interface fracture propagated. Special caution is needed during the initial steps of wedge advancement, because initially the vertical displacement only builds up stored energy without causing any fracture. Therefore the initial data points were discarded until it was clear that fracture had begun. The datum points in the final steps were discarded too. This is because in the final stage, the adhered part of the beam could be too short to be consistent with the assumption that the single beam is clamped which is required for the derivation of Eqs. 1 and 2 (Neglecting this, unusually large apparent **Gc** could be obtained).

### III Results

**1) Measured energy release rate**

The measured values of $G_C$ (for specimens processed in the same way) tended to display considerable scatter; not only from specimen to specimen, but also varying with the position of the crack as it propagated along the joint. Fig.2 shows $G_C$ as a function of crack length for four different TL specimens fired at 1150 ºC. It can be seen that the variability of **Gc** within a given specimen can be as large as that between specimens. The unusually low **Gc** values for one of the specimens (open circles) could have been due to the pre-existing large cracks or defects in the specimen generated during specimen preparation. Similar variability was found for all specimens prepared using different methods and fired at different temperatures.

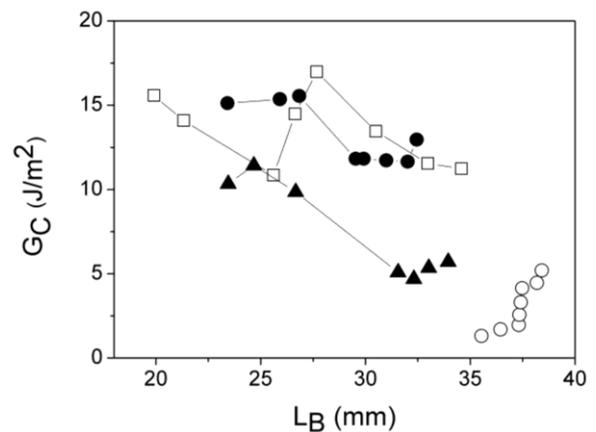

*Fig.2* $G_C$ *as a function of* $L_B$ *for different TL specimens all fired at 1150 ºC.*

Fig.3 shows $G_C$ as a function of firing temperature for specimens prepared using different methods. It can be seen that the $G_C$ for the SL specimens has no significant temperature dependence, while TL specimens show a large increase in $G_C$ when the firing temperature increases from 1100 to 1150 ºC. Furthermore, for the lower firing temperatures (1000 and 1100 ºC), the SL specimens have larger $G_C$ values than TL specimens, but the TL specimens have the larger $G_C$ value if fired at a higher temperature (1150 ºC).

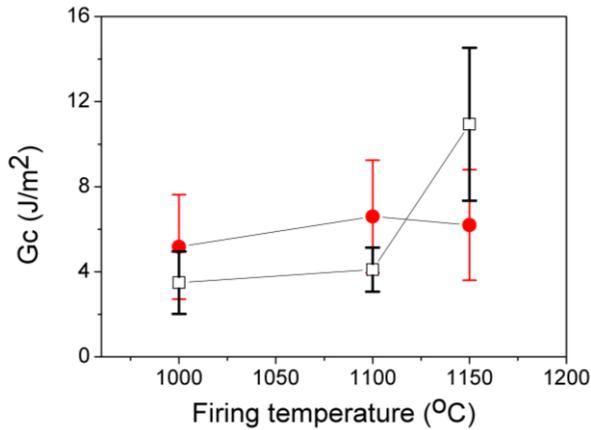

*Fig.3 $G_C$ as a function of firing temperature for specimens prepared using the SL method (solid circles) and the TL method (open squares). Each datum point was obtained by averaging the results from 4 specimens prepared using the same processing conditions and the error bars represent ± one standard deviation.*

**2) Fracture path**

The fracture path was characterised for its relative content of adhesive (at one of the LSCF/YSZ interfaces) or cohesive (within LSCF) failure. In most fractured specimen, it was possible to observe contributions from all three different fracture paths as the crack tip changed position within the joint as the crack advanced. Fig. 4 illustrates how the fracture path had deviated in post-test examination of a partially fractured specimen. The specimen was vacuum impregnated with low viscosity epoxy resin. After curing the specimen was cut parallel to the crack propagation direction and the cross section was polished. The result is shown in Fig. 4 in which fracture propagated from left to right. Fig.4a shows a main crack (adhesive delamination) at the lower substrate/LSCF interface, but there is some subsidiary damage at the upper beam/LSCF interface and a crack traversing the LSCF layer is also evident (arrowed). The field of view in Fig.4b is a short distance to the right (further along the fracture path) of Fig. 4a. Here the fracture at the upper interface is more apparent and a second crack traversing the LSCF is seen (arrowed). Fig.4c is further to the right, and here the crack at the lower interface is no longer evident, while the crack at the upper interface has become well-established. Thus the fracture path has crossed from one interface to the opposite one.

The location of the fracture path was also examined using optical microscopy in which YSZ appears white and LSCF is black. This strong contrast enables the fracture path to be visualized in plan view.

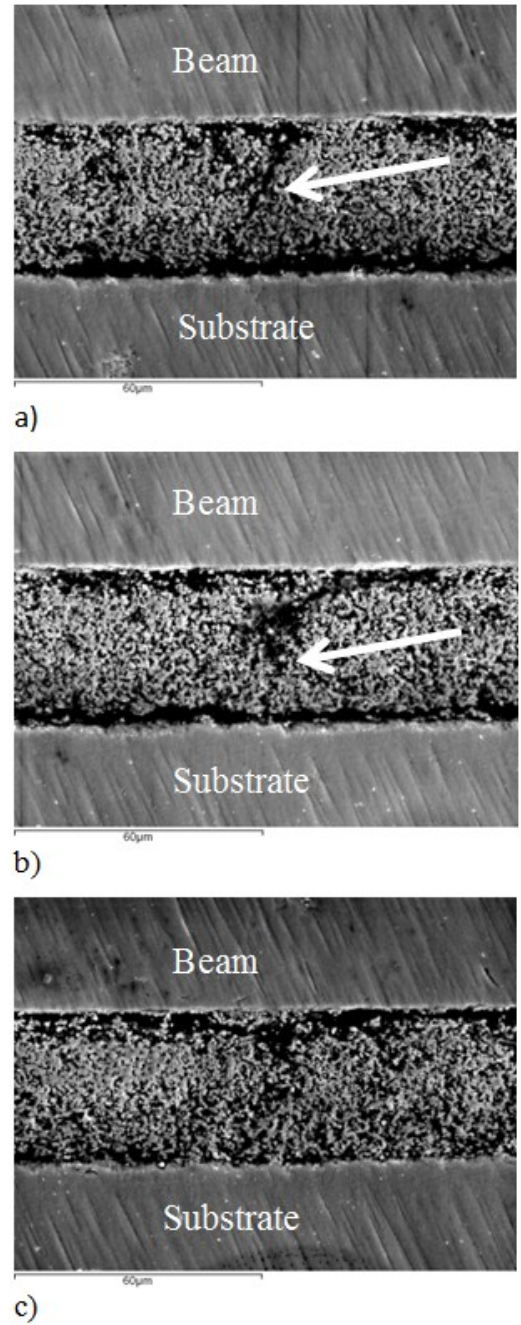

*Fig.4 Cross section of a partially failed specimen showing how the fracture path changed as the fracture propagated from left to right. a) a position near to the fracture initiation point, b) a short distance (a few millimeters) along the fracture; c) a further few millimeters along the fracture.*

Fig. 5 shows plan view pictures of the fracture faces of separated beam and substrate pairs for different specimens. The joint in Fig.5a (one of TL specimens fired at 1000 °C) has failed cohesively in the LSCF as both fracture faces (substrate and beam) are covered with thick residual LSCF.

The joint in Fig. 5b (a SL 1000 °C specimen) shows a mixture of cohesive (dark on both fracture faces) and adhesive failure (dark on one face matching light on the opposite one). It also shows some areas with no residual LSCF on both fracture faces. These are either regions in which adhesive failure has occurred at both interfaces (spalling) or, more probably, regions in which voids transversed a discontinuous LSCF layer (i.e. a fabrication defect). Fig.5c (TL 1150 °C specimen) shows a predominant adhesive failure at the upper YSZ/LSCF interface (the beam side), because not much residual LSCF can be seen on beam side. Fig.5d (SL 1150 °C specimen) also shows adhesive fracture from beam side interface, but also a significant fraction of areas that have no residual LSCF on both matching faces.

Quantitative image analysis using the software ImageJ was employed to determine the area fractions of different fracture modes on a given specimen. The optical images were first binarised using a threshold brightness that was checked manually to give accurate phase differentiation. As an example, Fig.6 a) and b) are the binarised images corresponding to Fig.5 d and show the distribution of YSZ (bright) and LSCF (dark) on the beam and substrate sides of the fracture path (The substrate side image, Fig.6 b, has been flipped vertically in order to match beam side image, Fig. 6a.) Simple image correlations were used to generate information about the fracture path. For example, areas with white pixels in Fig.6a corresponding to black pixels in Fig.6b show that these areas have adhesive failure on the beam side YSZ/LSCF interface. These areas are shown by bright in Fig.7a. Conversely, white areas in Fig.6b which correspond to dark areas in Fig.6a are areas with adhesive fracture on substrate side YSZ/LSCF interface and are shown bright in Fig.7b. The areas which are dark in both Fig. 6a and Fig. 6b are cohesive failures in the LSCF (bright in Fig.7c), while the areas which are bright in both Fig. 6a and Fig. 6b are areas in which both YSZ/LSCF interfaces failed or were voids (Fig.7d). In this way it was possible to quantify the area fraction of each joint that failed adhesively, $A_{ad}$, or cohesively, $A_{co}$.

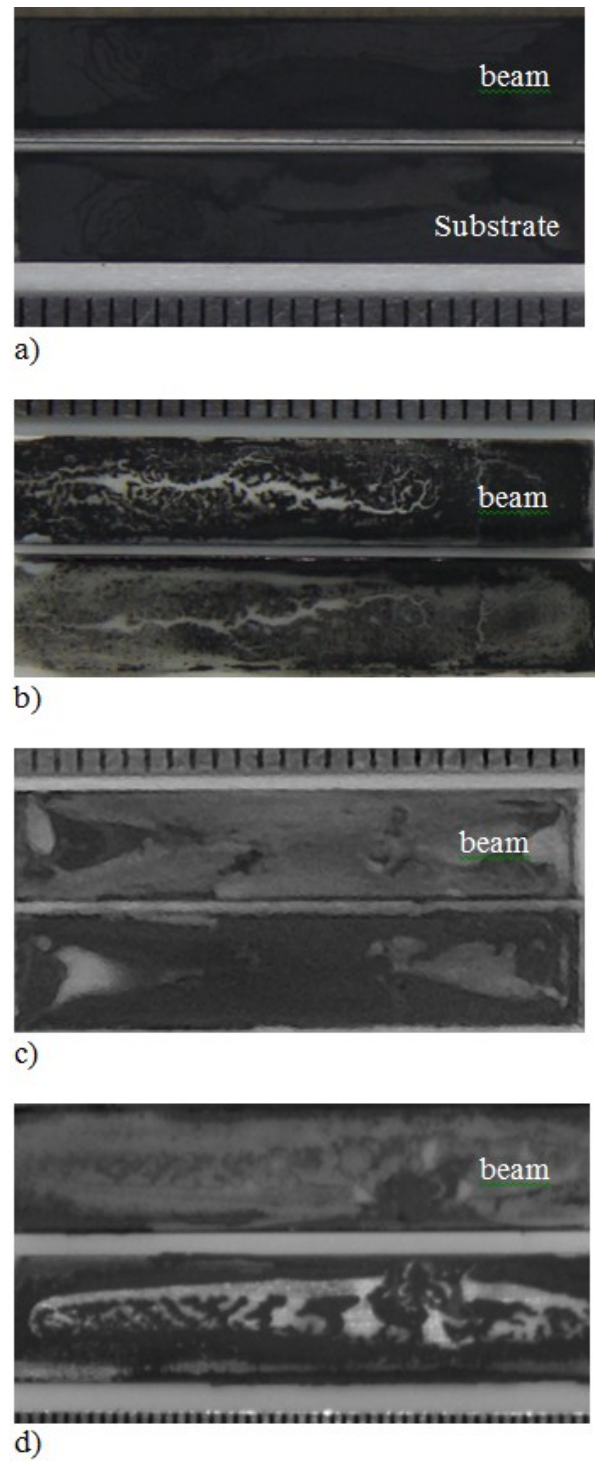

*Fig.5 Optical images of the separated (beam and substrate) pairs after the fracture test for a) a TL specimen sintered at 1000 °C, b) a SL specimen sintered at 1000 °C; c) a TL specimen sintered at 1150 C, and d) a SL specimen sintered at 1150 °C.*

Light areas are YSZ and dark areas LSCF. The pairs are oriented in mirror configuration. The graduation mark is in mm.

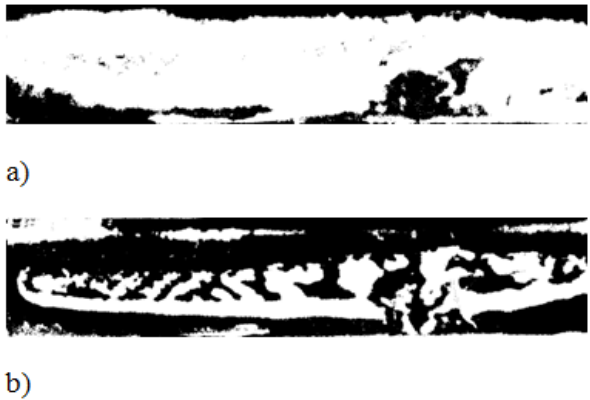

Fig.6 Binarised images of Fig. 5d showing the distribution of YSZ (white) and LSCF (black) on the beam and substrate side, respectively, of the fractured joint of a SL specimen sintered at 1150 °C

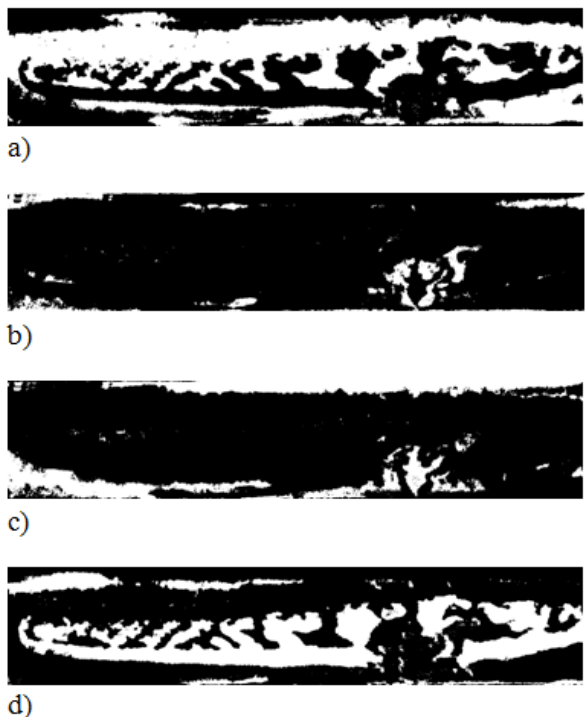

Fig.7 Binarised images from Fig. 5d and Fig. 6 showing in white: a) the areas with adhesive fracture along beam/LSCF interface; b) the areas with adhesive fracture along substrate/LSCF interface; c) the areas with cohesive fracture in the LSCF; d) the areas which are empty (voids). This particular example failed mainly adhesively at the beam/LSCF interface, but the joint had a large content of voids.

Fig. 8a and Fig. 8b summarise the image analysis results. As shown in Fig.8a, the total adhesive fracture area $A_{ad}$ for SL specimens does not seem to depend on the firing temperature, while it increases significantly with firing temperature for TL specimens. Consistent with this, in Fig. 8b the cohesive fracture area $A_{co}$ for SL specimens shows negligible dependence on firing temperature, while that for TL specimens shows a sharp decrease with the firing temperature.

For all the specimens the analysis showed that approximately 75% of the area of adhesive fracture was along beam/LSCF interface and 25% along the substrate/LSCF interface. This indicates beam/LSCF interface is more likely to fail than the substrate/LSCF interface in this single beam wedge test.

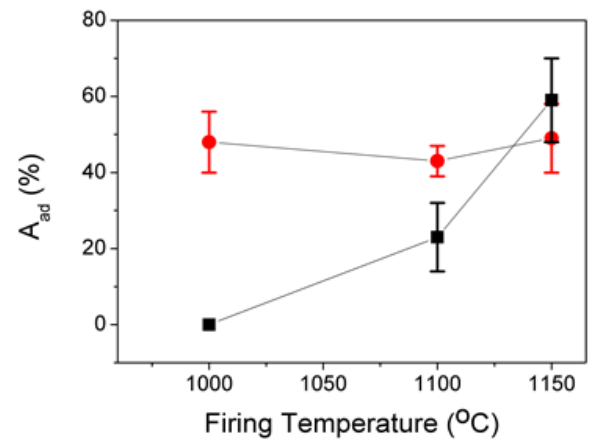

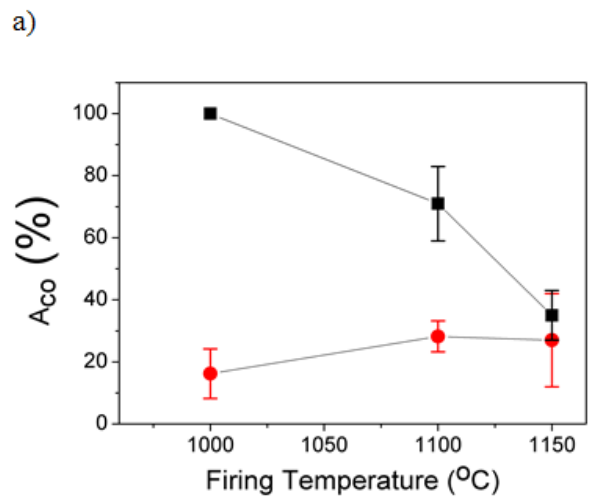

Fig. 8 a) Area fraction of adhesive fracture for SL specimens (solid circles) and TL specimens (solid squares) b) Area fraction of cohesive fracture for SL specimens (solid circles) and TL specimens (solid squares) fired at three different temperatures.

## IV Discussion

The single beam wedge test is a stable method: the propagation of interface crack reduces the stored elastic energy until the interface fracture resistance (toughness) is equal to or larger than the stored energy release rate (driving force). Due to the asymmetric geometry, the two YSZ/LSCF interfaces (i.e. beam/LSCF and substrate/LSCF interfaces) are not under identical stress conditions. To analyse the detailed stress conditions to which the two different interfaces are subjected during the test, finite element modelling (FEM) was performed (for details see the appendix). Fig.9a shows the maximum principle stress across the LSCF layer thickness (the origin corresponds to the substrate/LSCF interface and the maximum thickness value corresponds to the beam/LSCF interface). It is clear that the beam/LSCF interface is always subjected to a significantly higher stress than the substrate/LSCF interface, regardless the thickness of LSCF layer. Fig.9b shows the corresponding stress color map. This explains why beam-interface was observed to be more likely to fail than the substrate/LSCF interface.

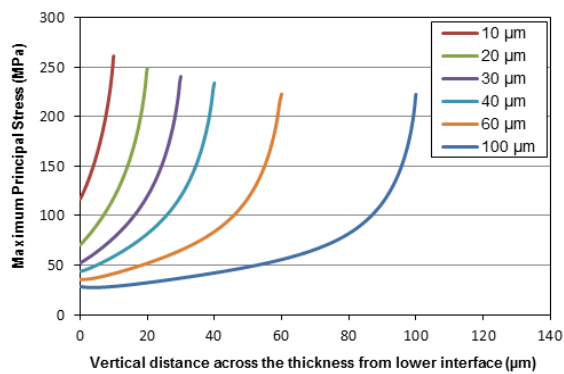

a)

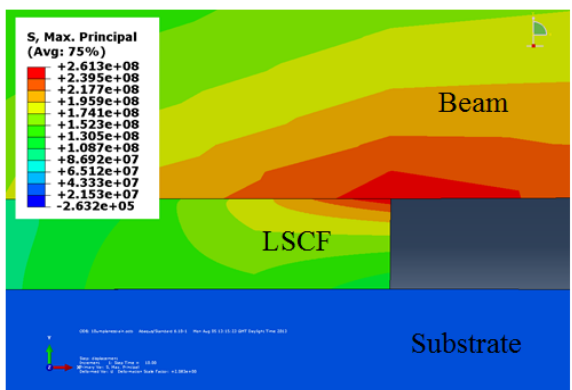

b)

*Fig.9 a) The maximum principle stress as a function of the distance from the LSCF/substrate interface at the free face of the LSCF layer for cases with different LSCF thickness; b) stress color mapping near the LSCF free face ( LSCF thickness=10μm). The loading conditions are given in the Appendix.*

The advantage of this single beam wedge test is that the joint fracture toughness at different steps (or locations along the crack propagation direction) can be obtained from a single specimen. The large variability in $G_C$ for the same specimens shown in Fig. 2 is not surprising as the porous microstructure of LSCF layer and variability of contact at the interfaces could vary significantly along the beam length. The large variability in interface microstructure and fracture locus along the beam length can be seen in Figs. 5b-d.

The LSCF layer in SL specimens was thin (about 10 μm) and had many voids (Fig.7d). The voids were generated during joint fabrication by confined spreading and drying of the ink. The area fraction of the voids in SL specimens was estimated from the image analysis to be 25-35%. In other words the effective contact area of the porous LSCF layer with YSZ was only 65-75% and this low value is an artifact of the SL fabrication method.

In contrast, TL specimens had no such voids. Image analysis revealed some TL specimens had <6% empty area which could be due to the double side fracture in some local areas (i.e. material spalls off both substrate and beam side). Despite the fact that TL specimens had much larger contact area at interfaces, the interface fracture toughness of TL specimens was not always larger. Under some conditions (i.e. firing temperatures ≤ 1100 ºC) TL specimens had even lower interface toughness than SL specimens (as shown in Fig.3).

To understand the measured $G_c$ it is necessary to look into the fracture path. The fracture path of SL specimens does not show significant dependence on firing temperature (solid circles in Figs.8a and b), while the fracture path of TL specimens shows a strong dependence (solid squares in Figs.8a and b). This suggests that the measured $G_c$ can be related to the fracture path. Figs.10a and b plot $G_c$ against adhesive fracture area $A_{ad}$ and cohesive area $A_{co}$ for TL and SL specimens respectively. It is clear from Fig.10a that the measured toughness of TL specimens increases with the adhesive fracture area and decreases with the cohesive fracture area. However, for SL specimens, the interface toughness is not strongly related to the fracture path (Fig.10b).

The bonding of the beam to the substrate by the porous LSCF is afforded by a network of particle 'chains' with one end connected to the beam and the other to the substrate. The inter-particle bonding would increase with firing temperature (due to sintering), whereas the bonding of the LSCF particles to YSZ depends not only on the firing temperature but also wettability between LSCF and YSZ. When the firing temperature is increased, the inter-particle bonding will be strengthened, but the bonding between LSCF particles to the YSZ can be limited by the wettability. TL specimens were approximately 3 times thicker than SL specimens. It is not surprising that the TL specimens would be more likely to experience cohesive fracture because long particle chains would have higher probability of having weakest link than short chains. This is especially so for the specimens fired at a relatively low temperature when inter-particle necking is relatively weak. At a sufficiently high firing temperature, the inter-particle bonding is better established and therefore cohesive fracture is more difficult. Thus a higher firing temperature led to more adhesive fracture and less cohesive fracture in TL specimens as shown in Fig.8.

For the SL specimens, the LSCF layer is thin and, as shown in Fig.8a and b, the cohesive fracture area is only about half that of adhesive fracture. In addition, the fracture path of the SL specimens did not show the significant dependence on firing temperature that was shown by TL specimens. This is probably because sintering of the very thin sandwiched layer is under more constraint than the thick sandwiched layer. For the sintering of the thick sandwiched layers in TL specimens, shrinkage normal to the substrate can be regarded as being free of constraint. But for the thinner SL layers, this is less likely because in some areas there is a high probability of larger particles, or particle agglomerates, spanning the layer thickness. In order to estimate the 'intrinsic' fracture toughness of the porous LSCF/YSZ interface ($G_{ad}$), it is necessary to remove the contribution to the measured fracture toughness of the joint from cohesive fracture. We can take the measured fracture toughness of the TL specimen fired at 1000 ºC, which showed 100% cohesive fracture, as a lower bound for cohesive fracture toughness ($G_{co}$). It is a lower bound because for higher firing temperatures the LSCF will have a higher cohesive fracture energy. Based on the equation: $G_{ad}A_{ad}+G_{co}A_{co}=G_c$, $G_{ad}$ values for the SL specimens and TL specimens are calculated and listed in Table 1.

Table 1 $G_{ad}$ for different specimens ($J\,m^{-2}$)

| Temperature | SL specimens | TL specimens |
|---|---|---|
| 1000 ºC | 9.6±3.5 | 9.6±3 |
| 1100 ºC | 13±3.9 | 7±3 |
| 1150 ºC | 11±4 | 16±5 |

Since the cohesive fracture toughness will not be a constant as assumed, the calculated values of $G_{ad}$ in Table 1 are only estimates. For example, the TL 1150 specimen appears to have a much higher $G_{ad}$ than the others, but this probably is a consequence of taking the lower bound for $G_{co}$ independent of firing temperature. Nevertheless, the results in Table I are sufficient to show that a porous LSCF/dense YSZ interface typical of a fuel cell cathode has a fracture energy of approximately 11 ± 2 $J\,m^{-2}$ and is relatively insensitive to firing temperature within this narrow range of 1000 – 1150 ºC. This interface fracture energy value is significantly larger than that of the interface between dense lanthanum strontium chromite (LSC) and a porous lanthanum strontium manganite (LSM) which was 1.4 –3.8 $J\,m^{-2}$ (measured by a double cantilever beam method [16]), but smaller than that of the interface between 3YSZ and porous LSM which was 20.2 $J\,m^{-2}$ (measured by modified 4-point bending method [3]). Compared to the toughness of the fully dense LSCF which is about 1.5 $MPa\,m^{1/2}$ (equivalent to 15 $J\,m^{-2}$), a fracture toughness of 11 $J\,m^{-2}$ for the porous LSCF/3YSZ interface is unexpectedly large (considering the porosity of the interface is large, maybe as high as 50%). The tetragonal phase in 3YSZ can be triggered mechanically to transform to monoclinic phase, which is accompanied by 5-7% volume expansion [23-25]. This would absorb significant amount of energy, leading to the toughening of YSZ containing materials [23, 25]. The phase transformation toughening mechanism might also lead to the toughening of porous material/3YSZ interfaces. This might explain why the porous LSCF/3YSZ in this work and porous LSM/3YSZ in [3] show unusually high interface fracture toughness. However, further works are needed to clarify this.

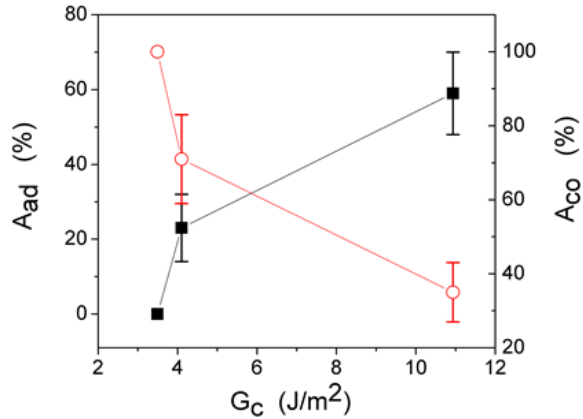

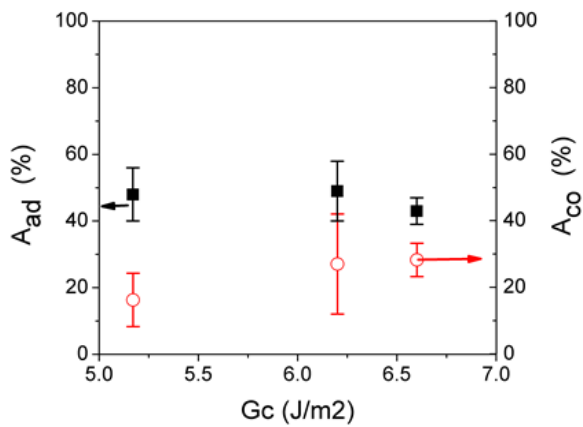

*Fig.10 The dependence of the measured value of $G_c$ on adhesive fracture area ($A_{ad}$) and cohesive fracture area ($A_{co}$) for a) TL specimens and b) SL specimens.*

## V Conclusions

1) Single cantilever beam wedging is a suitable method for measuring the fracture toughness of joints having a porous LSCF adhesive between dense YSZ adherands. The fracture in the joint proceeds stably and the crack tip position can be determined with acceptable precision from the bending profile of the cantilever beam.
2) The fracture mode of such joints was found to depend on both the firing temperature and the thickness of the porous LSCF layer. In particular a high firing temperature and small layer thickness led to more adhesive fracture and less cohesive fracture.
3) The measured joint toughness depended strongly on the fracture mode (adhesive fracture area relative to cohesive fracture area). A larger adhesive fracture area was related to a higher joint toughness.
4) The intrinsic adhesive fracture toughness for the porous LSCF/dense YSZ interface is estimated to be 11 J m$^{-2}$ for specimens fired at temperatures between 1000-1150 C. It is speculated that the reason the firing temperature had little influence on the interface toughness was possibly due to a restriction on the sintering between the LSCF particles and the YSZ substrate related to their interfacial energies.

## Appendix

Finite element modelling (FEM) of stresses in the single cantilever beam wedge test

The single cantilever beam wedge test was modelled in 2 dimensions, for which a schematic is shown in Fig.A1. The material properties used were: Young's modulus, $E$, = 200 GPa, Poisson's ratio, $v$, = 0.3 for dense bulk YSZ; and $E$ = 50 GPa, $v$ = 0.3 for the porous LSCF (regarded as a homogeneous material). The standard FE solver in Abaqus CAE 6.10 (Dassault Systemes, USA) was used for simulation. A 4-node bilinear plane stress quadrilateral element was applied to generate the simulation mesh. In order to yield accurate simulation results, a finer mesh was used in regions close to free face of the LSCF interlayer, as marked with the red oval in the schematic.

FEM was carried out based on the assumption that the parts were homogeneous, isotropic and linear elastic. The simulation was under 2D plane strain condition and the bottom edge of the YSZ substrate was fixed. A vertical displacement of 2.4 mm (typical for the actual experiments) was applied on the bottom node (labelled "A" in the schematic) on the right edge of the upper YSZ beam. The modeled results are shown in Fig.9.

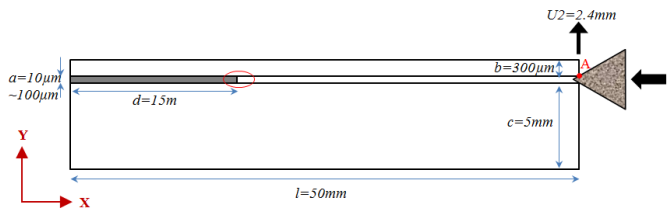

*Fig.A1 Schematic for FE modelling of the single cantilever beam wedge test.*

Fig.A1 is relevant to the initial state of the specimen. Two other cases of interest are 1) a crack propagates along the beam/LSCF interface as shown in Fig. A2; and 2) a crack propagates along the substrate/LSCF interface as shown in Fig. A3.

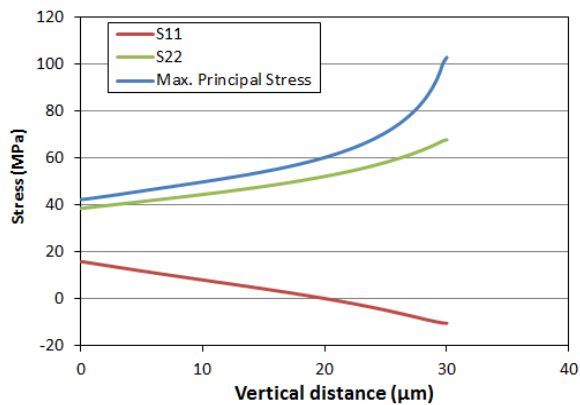

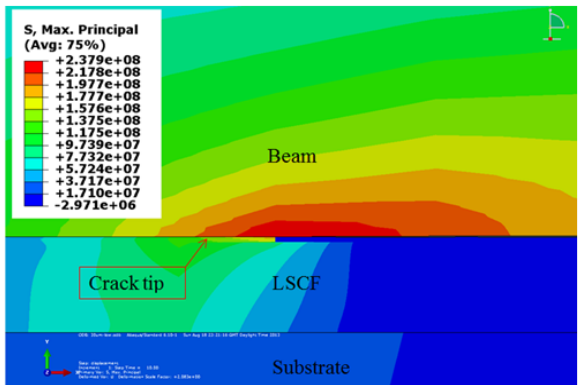

*Fig.A2 a) Stress as a function of the distance from substrate-LSCF interface at the free face of a 30μm thick LSCF layer loaded as in Fig. 1. b) Map of the normal stress component $S_{22}$ near a crack tip for the case where the delamination has initiated at beam/LSCF interface.*

As shown in Figs. A2 a and b, the beam/LSCF interface is subjected to higher stresses (both normal and maximum principal stress) than the substrate/LSCF interface. This implies that delamination would preferably propagate along the beam/LSCF interface if it initiates at the beam/LSCF interface.

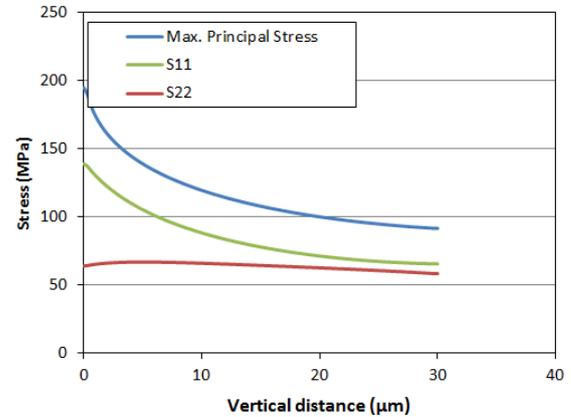

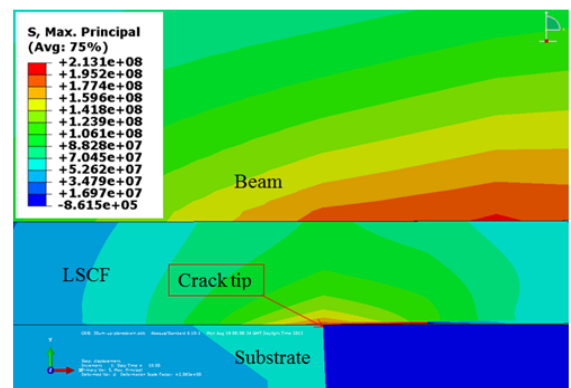

*Fig.A3 a) stress changes with the distance from substrate-LSCF interface for 30μm thick LSCF layer, b) stress color mapping near crack tip for the case where the delamination initiated at substrate-LSCF interface.*

If the initial delamination is along the substrate/LSCF interface (as shown in Fig.A3), the vertical stress component ($S_{22}$) remains almost constant cross the thickness for this case, but the lateral stress component ($S_{11}$, ~140 MPa at the substrate/LSCF interface) is much larger than $S_{22}$ (~65MPa) and decreases with the distance from the substrate/LSCF interface. The large lateral stress would very likely generate vertical cracking in LSCF layer. This implies the initial cracking along the substrate/LSCF interface is very likely to jump over to the top beam/LSCF interface.

Therefore overall, the substrate/LSCF delamination is a relatively less likely event.

**Acknowledgements**
This research was carried out as part of the UK Supergen consortium project on "Fuel Cells: Powering a Greener Future". The Energy Programme is an RCUK cross-council initiative led by EPSRC and contributed to by ESRC, NERC, BBSRC and STFC. Fei He thanks for the financial support from China Scholarship Council (CSC) (No. 2009307054) to carry out one year visiting research at Imperial College London.